\documentstyle[epsf]{article}
\begin{document}
\renewcommand{\thefootnote}{\fnsymbol{footnote}}
\begin{flushright}
KEK Preptint 93-43\\
June 1993
\end{flushright}
\vskip -2cm
\epsfysize3cm
\epsfbox{kekm.epsf}
\begin{center}
{\large \bf
Test of Various Photocathodes
\footnote{
published in Nucl. Instrm. Meth. {\bf A 343} (1994) 117.
}}\\
Ryoji Enomoto, Takayuki Sumiyoshi\\
{\it National Laboratory for High Energy Physics,
1-1 Oho, Tsukuba-shi, Ibaraki 305, Japan}\\
 and Yoshio Fujita\\
{\it Hamamatsu Photonics K. K., \\
314-5 Shimokanzo, Toyooka Vill., Iwata-gun, Shizuoka 438-01, Japan}\\
\end{center}
\begin{abstract}
A test of various photocathodes was carried out.
The tested materials were CsI, CsTe,
their multi-layers and so on. The quantum efficiencies of the various
materials were measured under a vacuum and/or after exposure to several
kinds of gases.
\end{abstract}

\section{Introduction}

Various attempts to understand the CP violation mechanism[1]
are presently being planned.
$B$ mesons seem to be a most promising
probes to be investigated[2]. In order to carry out
such investigations, B-factory projects have been proposed[3].
One of them is the KEK B-factory project,
in which asymmetric double storage rings to produce high-rate
moving-$B\bar{B}$s will be constructed[4].
To measure CP asymmetry, we need to identify the b-flavor: in other
words either $B$ or $\bar{B}$. The best probes have been found to be
charged $K$ mesons from the cascade decays of $B$ mesons[5].
$\pi$/$K$ separations in the momentum range between 0.5-4 GeV
are required.

In addition, a fast response of the detector is required,
due to the use of a high-luminosity accelerator. We have adopted a fast
ring imaging Cerenkov detector
(in short, ``fast RICH"[6])
for $\pi$/$K$ identification.
Solid photocathodes which can be
operated even after exposure to air have recently been
discovered; they were also proven to be useful in a fast
RICH[7]. The most stable (to air-exposure) photocathode was
found to be CsI coated with
a thin layer of TMAE.
However, the quantum efficiencies have not yet been established
in the presence of air or MWPC-gases[8].

We have systematically carried out measurements of
the quantum efficiencies of CsI
as well as other materials. The optimization
of a CsI photocathode and the search for other materials
have been carried out.

\section{Test Setup}
In order to measure the quantum efficiencies of various photo-sensitive
materials, we have adopted a phototube technique. A test phototube
(diode) is shown in Fig 1.
\begin{figure}
\vskip 3cm
\epsfysize5cm
\epsfbox{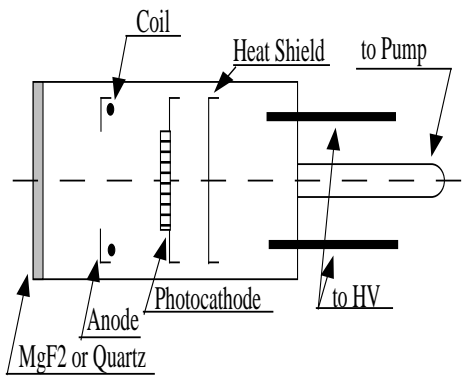}
\caption{Cross-sectional view of the test phototube. The cathode
was made of Al-plated Ni-electrode. The photocathodes were plated
on it.}
\end{figure}
Window were made of $MgF_2$ or
quartz. The cathode was made of an Al-plated-Ni electrode,
and various photo-sensitive materials were plated on it.
Tests were then carried out at room temperature.
The spectral responses were measured using monochrometers.

\section{Test on a Single Material in a Vacuum}

\subsection{CsI}
At first, we optimized the thickness of the CsI photocathode.
The quantum efficiencies were simultaneously
monitored at a wavelength of 170 nm
with vacuum-plating CsI. The result
is shown in Fig 2.
\begin{figure}
\vskip 3cm
\epsfysize5cm
\epsfbox{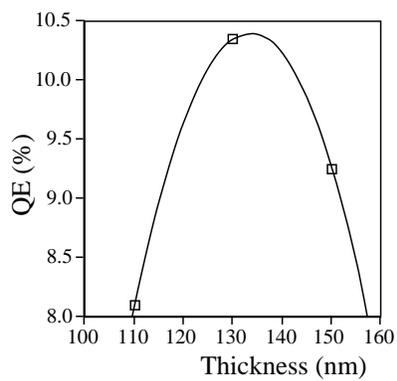}
\caption{Quantum efficiencies versus the thickness of the CsI
photocathode.}
\end{figure}
The real thickness of the CsI photocathode was calibrated, and the
correction factor was typically $\sim$30\%.
We therefore concluded that the optimized thickness was
approximately 130 nm.
This value is different from that of reference [7] about a factor
of 4. A factor of 2 can be explained by reflection at the
Al-surface of the electrode. There may be a systematic
difference in the CsI quantum efficiency due to such circumstances
as gas. Our measurement was carried out under a vacuum.
We replaced the cathode-electrode with an stainless steel which has
a higher reflectivity, however, the quantum efficiency was the same.

The spectral response of the quantum efficiency is shown in Fig 3.
\begin{figure}
\vskip 3cm
\epsfysize5cm
\epsfbox{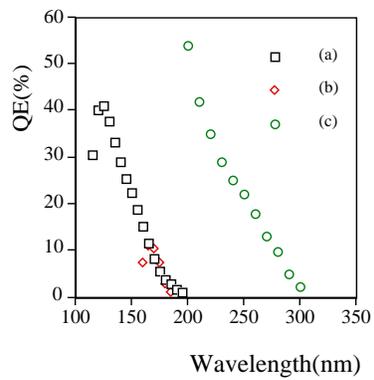}
\caption{Quantum efficiencies of optimized photocathodes: (a) CsI
with a MgF$_2$ window, (b) CsI with a quartz window, and (c) CsTe with
a quartz window.}
\end{figure}
The points indicated by (a) were measured using a $MgF_2$ window
and by (b) a quartz window. The peak is located at $\sim$120 nm (QE=40\%),
where $MgF_2$ cutoff is 115 nm.

\subsection{CsTe}
The CsTe photocathode has been known to be sensitive at longer
wavelengths and to have a higher QE than that of CsI.
This photocathode was tested with a quartz window phototube.
Although the optimized thickness was located
at around 100-300 nm, this requires
more study. The quantum efficiency was plotted
with the points indicated by (c)
in Fig 3. A QE of 55\% was obtained at 200 nm.

\subsection{CsI(Tl)}
We next tested a Tl-doped CsI
in order to find any difference in the spectral
response.
After that, $MgF_2$ windows were used in the test phototubes.
The measurements were carried out at the optimized thickness.
The measured quantum efficiency is shown in Fig 4.
\begin{figure}
\vskip 3cm
\epsfysize5cm
\epsfbox{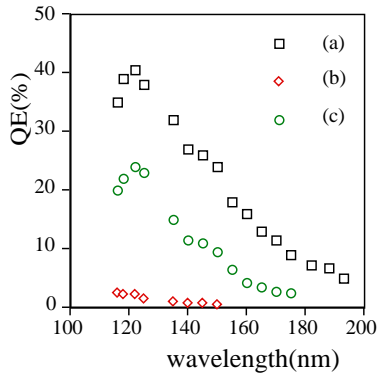}
\caption{Quantum efficiencies of various photocathodes:
(a) CsI(Tl), (b) NaI(Tl), and (c) KI.}
\end{figure}
We did not observe any spectral differences, compared
with those found in the pure CsI.
This photocathode was thus also proven to be stable.

\subsection{NaI(Tl)}
NaI is known to have a lower work function (Eg), and slightly
higher electron affinity (Ea).
We anticipate the shift of the sensitivity at the longer wavelength
region. The result is shown in Fig 4.
Unfortunately, the quantum efficiency is
too low to be used in our case.

\subsection{KI}
KI has a similar Eg and Ea compared to that of NaI. The result
is shown in Fig 4. A quantum efficiency of 24\% at $\sim$120 nm
was obtained. However, the sensitivity in the longer wavelength
region
was significantly lower than
that of CsI. There is no effective sensitivity with
a quartz window.

\section{Quantum efficiency after gas exposure}
In this chapter we report on the sensitivity changes
occurring after exposing
photosensitive materials to various gases.

\subsection{CsI}
At first we tried $N_2$. After 1 hour of exposure to $N_2$, the quantum
efficiency was reduced by about 5\%.
In these measurements, the thickness of CsI was
130 nm. The dependence on the thickness will be measured
in the near future.
We next tried $C_3H_8$, and a 50\% reduction of the quantum efficiency was
observed.
$NH_3$ was also tried. The sensitivity was found to be stable against
exposure. CsI was found to be stable against amin-gas.

We finally we tried air exposure.
The result is shown in Fig 5.
\begin{figure}
\vskip 3cm
\epsfysize5cm
\epsfbox{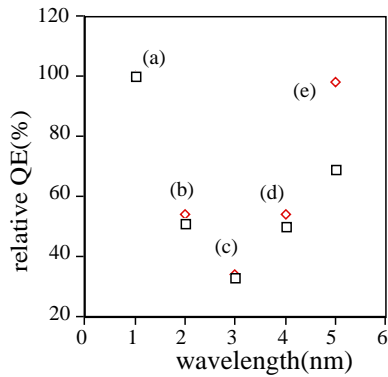}
\caption{Relative quantum efficiencies after exposure to the air:
(a) before exposure, (b) after 1 hour of exposure, (c) after 2 hours
of exposure, (d) after 3 hours of exposure, and (e) 5 days after (d)
in a vacuum. The
squares are the relative QE at 121.6 nm, and the circles are at
160.8 nm.}
\end{figure}
We monitored the anode current at two wavelengths of
light sources (121.6
and 160.8 nm). There were 5 steps: (a) before exposure, (b)
after 1 hour of exposure, (c) 2 hours, (d) 3 hours, and
(e) leaving the
test photocathode for 5 days in vacuum after (d).
Although the quantum efficiencies were reduced by the exposures,
they recovered after more than 20 hours of vacuum pumping.

We tried yet another test. There was one exposure of 4 hours; after that,
the test photocathode was left in a vacuum. The results are shown in Fig 6.
\begin{figure}
\vskip 3cm
\epsfysize5cm
\epsfbox{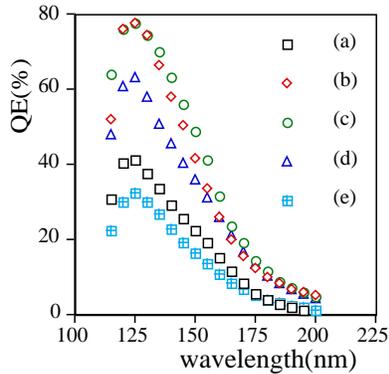}
\caption{Time variation of the quantum efficiency of the CsI photocathode
after exposure to air: (a) before exposure,
(b) after 4 hours of air-exposure,
(c) after leaving in a vacuum for 17 hours, (d) after 3 days,
and (e) after 8 days.}
\end{figure}
There was a drastic change in the sensitivity. The quantum
efficiency became as high as 77\% at $\sim$120 nm (before exposure,
it was only 40\%).
However, it was not stable after the vacuum-pumping. There was
a possibility that some contamination in the air contributed to the
increase in
the quantum efficiencies of CsI. We need
to carry out a more systematic check
of this effect.

\subsection{CsTe}
CsTe showed a similar result
compared to that of CsI regarding $N_2$ exposure.
A 5\% reduction of the quantum efficiency was observed.
Air exposure was serious, and the quantum efficiency dropped
to less than 1\%.

\section{Double-layer photocathodes}
CsI was proven to be stable against air exposure.
We thus tried to coat an unstable material with CsI (a double
photocathode).

\subsection{CsTe + CsI}
The most promising UV photocathode was CsTe. It has a higher
sensitivity at wavelengths greater than 200 nm. The advantage involves
the use of a quartz window. We tried to coat CsTe with CsI.
Unfortunately the result was negative. After plating a thin film of
CsI, CsTe itself reacted with CsI and the quantum efficiency dropped
rapidly, even in a vacuum. After air exposure, additional
damage was observed.

\subsection{CsSb + CsI}
CsSb was known to match CsI in a vacuum,
and it shows a good sensitivity to visible light.
We tried to coat a CsSb photocathode with CsI [9].
At first in vacuum, the coating was carried out by a few nm steps.
After each step, the quantum efficiency was observed to be reduced by
10-20\%. After 6 steps, we exposed it to air and
then measured the quantum efficiency.
It had become reduced by one order (i.e., less than 1\%).
We therefore concluded
that we cannot prevent a reaction with air by using a thin film of
CsI.

\subsection{CsI + NaI}
NaI was stable in air. We thus tried to coat NaI with CsI and CsI with NaI.
This was a miscellaneous test in order to understand any changes
in the quantum efficiencies by a coating procedure. In the first
case, the quantum efficiency at 170 nm was the same as
that of pure CsI,
and that at 180 nm was half. In the latter case,
however, the quantum efficiency
at 170 nm was half and at 180 nm it was 1/4.
There was a reduction by the double-layer technique.
\section{Conclusion}
We tested various photocathodes
for gas-filled photon detectors, such as a ring-imaging Cerenkov
detector using test phototubes.
Various materials which were stable to air exposure were tested.
Among them, CsI was found to be
the most suitable, and various optimizations
were carried out. Other attempts, such as
using new materials other than CsI, or
using double-layer photocathodes did not show any stable results.

\section*{Acknowledgement}
We thank a discussion with Prof. T. Ypsilantis (College de France).
We appreciate the support given by Prof. F. Takasaki in carrying out
this R and D. This work was carried out under collaboration with
Hamamatsu Photonics K. K. (HPKK). Especially, the work by Mr. T. Hakamada,
Mr. S. Suzuki, and Mr. Y. Hotta is acknowledged.

\noindent{\large \ {REFERENCES:}}

\noindent{[1] M. Kobayashi and T. Maskawa, Prog. Theor. Phys., 49 (1973)
652.}
\newline
\noindent{[2] I. I. Bigi and A. I. Sanda, Nucl. Phys. B193,
(1981) 85.
}
\newline
\noindent{[3] P. J. Oddone, private communication.
}
\newline
\noindent{[4] KEK B-factory Task Force, KEK report 90-23, unpublished.
}
\newline
\noindent{[5] R. Aleksan et. al., Phys. Rev. D39 (1989) 1283.
}
\newline
\noindent{[6] R. Arnold et. al., Nucl. Instrum. Methods, A314 (1990) 465.
}
\newline
\noindent{[7] J. Seguinot et. al., Nucl. Instrum. Methods, A297 (1990) 133.
}
\newline
\noindent{[8] B. Howeneisen et. al., Nucl. Instrum. Methods, A302 (1991) 447.
}
\newline
\noindent{[9] suggested by T. Ypsilantis (College de France).}

\end{document}